\documentclass[onecolumn]{IEEEtran}
\usepackage{times}
\usepackage{cite}
\usepackage{color}
\usepackage{epsf}
\usepackage{epsfig}
\usepackage{graphicx}
\usepackage{graphics}
\usepackage{epstopdf}
\usepackage[small]{caption}
\usepackage{hyperref}
\usepackage{amsmath}
\usepackage{amssymb}
\usepackage{amsthm}
\usepackage{amsxtra}
\usepackage{multirow}

\usepackage[ruled]{algorithm2e}
\usepackage{algorithmic}
\usepackage{enumerate}
\usepackage{multirow}
\usepackage{enumerate}
\usepackage{amssymb}
\usepackage{lipsum}
\usepackage{setspace}
\usepackage{multirow}
\usepackage{wasysym}
\usepackage{enumerate}
\usepackage{amssymb}
\usepackage{lipsum}
\usepackage{setspace}
\usepackage{multirow}
\usepackage{subcaption}
\usepackage{array}
\usepackage{makecell}
\usepackage[first=0,last=9]{lcg}

\definecolor{Gray}{gray}{0.9}

\newtheorem{definition}{\bf Definition}
\newtheorem{remark}{Remark}

\usepackage{caption} 
\usepackage{rotating} 
\usepackage{graphicx} 
\usepackage{colortbl} 

\newcommand{\Rmnum}[1]{\expandafter\@slowromancap\romannumeral #1@}

\makeatother
\begin{document}
\title{A Motivational Game-Theoretic Approach for Peer-to-Peer Energy Trading in the Smart Grid}
\author{Wayes Tushar$^{1,*}$, Tapan Kumar Saha$^1$, Chau Yuen$^2$, Thomas Morstyn$^3$, Malcolm D. McCulloch$^3$, H. Vincent Poor$^4$, and Kristin L. Wood$^2$\\$^1${The University of Queensland, Brisbane, Australia}\\$^2${Singapore University of Technology and Design, Singapore}\\$^3${The University of Oxford, Oxford, United Kingdom}\\$^4${Princeton University, NJ, USA}
\thanks{$^*$Corresponding author: School of Information Technology and Electrical Engineering, The University of Queensland, Brisbane, QLD 4072, Australia.}
\thanks{\emph{Email address:} w.tushar@uq.edu.au (W. Tushar), saha@itee.uq.edu.au (T. K. Saha), yuenchau@sutd.edu.sg (C. Yuen), thomas.morstyn@eng.ox.ac.uk (T. Morstyn), malcolm.mcculloch@eng.ox.ac.uk (M. D. McCulloch), poor@princeton.edu (H. V. Poor), kristinwood@sutd.edu.sg (K. L. Wood).}}
\IEEEoverridecommandlockouts
\maketitle
\doublespace
\begin{abstract}
Peer-to-peer trading in energy networks is expected to be exclusively conducted by the prosumers of the network with negligible influence from the grid. This raises the critical question: \emph{how can enough prosumers be encouraged to participate in peer-to-peer trading so as to make its operation sustainable and beneficial to the overall electricity network?} To this end, this paper proposes how a motivational psychology framework can be used effectively to design peer-to-peer energy trading to increase user participation. To do so, first, the state-of-the-art of peer-to-peer energy trading literature is discussed by following a systematic classification, and gaps in existing studies are identified. Second, a motivation psychology framework is introduced, which consists of a number of motivational models that a prosumer needs to satisfy before being convinced to participate in energy trading. Third, a game-theoretic peer-to-peer energy trading scheme is developed, its relevant properties are studied, and it is shown that the coalition among different prosumers is a stable coalition. Fourth, through numerical case studies, it is shown that the proposed model can reduce carbon emissions by 18.38\% and 9.82\% in a single day in Summer and Winter respectively compared to a feed-in-tariff scheme.  The proposed scheme is also shown to reduce the cost of energy up to 118 \cent~  and 87 \cent~ per day in Summer and Winter respectively. Finally, how the outcomes of the scheme satisfy all the motivational psychology models is discussed, which subsequently shows its potential to attract users to participate in energy trading.
\end{abstract}
\section*{Nomenclature}
\addcontentsline{toc}{section}{Nomenclature}
\begin{IEEEdescription}[\IEEEsetlabelwidth{$V_1,~~V_2,$}]
\item[$n$] Index of each prosumer.
\item[$E_{n,s}$] Energy surplus of prosumer $n$.
\item[$E_{n,d}$] Energy deficiency of prosumer $n$.
\item[$\mathcal{N}_s$] The set of sellers of energy.
\item[$\mathcal{N}_b$] The set of buyers of energy.
\item[$\nu$] The value of the coalition.
\item[$p_{g,s}$] Selling price of the grid.
\item[$p_{g,b}$] Buying price of the grid.
\item[$E_s$] Difference between total surplus and total demand in the network.
\item[$E_d$] Negative value of $E_s$.
\item[$p_s$] Peer-to-peer selling price.
\item[$p_b$] Peer-to-peer buying price.
\item[CO$_2$] Carbon dioxide.
\item[DER] Distributed energy resource.
\item[FiT] Feed-in Tariff.
\item[P2P] Peer-to-peer.
\item[ISO] Independent system operator.
\item [CPS] Central power station.
\end{IEEEdescription}
 \setcounter{page}{1}
\section{Introduction}\label{sec:introduction}
Global warming and the resultant climate change has become an important global issue recently, and using environmentally friendly technologies for the energy sector is recognized as a potential solution to this problem \cite{MoJang-Elsevier_2015}. As such, there has been a significant increase in adopting distributed energy resources (DERs) at the edge of the grid level~\cite{Jin_AE_Jan_2018}, i.e., at the consumer premises. For example, the global market for rooftop solar panels is forecast to be increased by $11\%$ compared to its market value in $2016$ over the next five years. This increase will further be complemented by a rise of $3605$ MW in the supply of electricity from residential storage devices by $2025$ \cite{Peck_Spectrum_2017}. 

The expected benefit of such widespread deployment of DERs, however, is contingent on the extensive participation of the owners of these assets, i.e., the prosumers \cite{Morstyn_NE_2018} in the electricity market. For this purpose, a number of strategies are being used at present to engage DER owners into the electricity market. Examples of such strategies include direct method~\cite{Ruiz_TPS_2009}, indirect method \cite{Morstyn_NE_2018}, and feed-in-tariff (FiT) \cite{Liang-ChengYe-AE:2017} scheme. In the direct method, individual DER is controlled and managed by an aggregator, whereas indirect methods refer to strategies in which a central authority sends a signal to the owners to influence their consumption and generation decisions. Under a FiT, however, the engagement of prosumers is more unequivocal. That is, the prosumers sell their excess electricity directly to the grid and buy electricity from the grid, if needed. 

While different strategies have their own advantages depending on the application, the benefit to prosumers needs to be fully realized for their sustainable engagement in the energy market. In other words, energy management techniques need to be prosumer-centric and to encourage the extensive participation of prosumers by demonstrating the benefits of participation~\cite{Tushar_TSG_2014}. Otherwise, prosumers may decide to go off-grid bringing inefficient outcomes to both the grid and the prosumers \cite{Morstyn_NE_2018}. Examples of such cases can be found in \cite{Colley_Misc_2014} (failure of smart meter trial, which is used in direct and indirect methods) and \cite{FiTPolicy_2015} (discontinuation of FiT).

Peer-to-peer (P2P) trading~\cite{Zhang_AE_June_2018} has emerged as a new energy management paradigm that allows energy from one prosumer to go to another prosumer without any direct influence of a central controller \cite{Tushar_SPM_2018} and is expected to address the limitation of existing techniques. In a P2P trading platform, a prosumer with the need of energy can take advantage of other prosumers within its community with energy surplus by buying the excess energy at a relatively cheaper rate \cite{Tushar_TIE_2015}. On the other hand, prosumers with excess energy can make more revenue compared to the FiT scheme by participating in P2P energy trading. Due to the potential benefits, a large number of studies are conducted to explore the potential of P2P energy trading among different prosumers. Details of these existing studies are provided in Section~\ref{sec:preliminaries}. Further, a number of pilot projects are currently being developed across the world, including the Brooklyn Microgrid in New York \cite{Mengelkamp_AppliedEnergy_2017}, Valley Housing Project in Fremantle, Western Australia \cite{Anant_thesis:2017}, and the Enerchain Project in Europe \cite{enerchain_Misc_2017}.

Most existing studies have contributed to the problem of devising energy trading algorithms~\cite{Sorin_IEEE_2018}, modeling the trading price~\cite{Liu_TPS_2017}, incentivizing prosumers~\cite{Han_IEEE_2019}, maintaining network constraints~\cite{Chapman_IEEE_2018}, and designing business contracts between different prosumers~\cite{Morstyn_TSG_2018}. However, a key gap which has not been addressed is: \emph{how to design a prosumer-centric energy trading between different peers\footnote{Hereinafter, peer and prosumer will be used interchangebly to refer to a prosumer in the manuscript.} within a community that will ensure their widespread and sustainable engagement in the electricity market?} This issue is particularly important due to the fact that P2P trading platform is a trustless system without the presence of a central controller (such as a retailer) for coordinating the energy trading \cite{PowerLedger_2017}. Therefore, \emph{encouraging prosumers to trust and cooperate} to trade their energy with one another could be significantly challenging.

To this end, the main innovation of this study lies in exploring how motivational psychology ~\cite{Kuwabara_AMP_2015} can be exploited for designing a P2P energy trading technique in order to bridge the gap between the research and development of P2P energy trading by providing an insight into how prosumers can be motivated to participate. In doing so, we have made the following contributions in this paper:
\begin{itemize}
\item A detailed overview of existing literature on P2P energy trading is provided following a systematic classification. This would help the reader to understand the current state of research in P2P trading and how the proposed study contributes to the relevant body of knowledge.
\item Motivational psychology is identified as a new tool for designing energy trading. The paper provides a detailed introduction to the concept, and shows how it can be exploited through its various motivational models to design P2P energy trading schemes with a view to encourage extensive prosumer participation.
\item A demonstration of the design process of a P2P energy trading scheme is provided using cooperative game theory. The properties of the game are studied and the stability of the developed scheme is validated. Then, by comparing the properties of the outcomes with that of motivational models, it is shown that the developed scheme satisfies all the properties of the discussed motivational models and thus has the potential to enable widespread prosumer participation in the P2P trading.
\item Finally, some numerical case studies are provided to corroborate the claim about the developed P2P trading scheme that it satisfies the necessary motivational models.
\end{itemize}

The rest of this paper is organized as follows. In Section \ref{sec:preliminaries}, the terms P2P trading and motivational psychology that are extensively used in this paper are introduced. How the P2P trading can be designed through a motivational psychology based coalition game is explained in Section \ref{sec:gamedesign}. Finally, some concluding remarks are drawn in Section \ref{sec:conclusion}.

\section{Preliminaries}\label{sec:preliminaries}In this section, a detailed overview of  P2P energy network and motivational psychology is provided. The objective is to present the readers with a clear understanding of how they could potentially be integrated together for designing a sustainable prosumer-centric P2P energy trading scheme in the future.
\subsection{P2P Energy Network}According to \cite{Schollmeier_2001}, a distributed network architecture can be called a P2P network, if the participants of the network share a part of their own resources with one another. These shared resources are necessary to provide the service and content offered by the network (for example, file sharing) and can be accessed by other peers directly, without passing intermediary entities. Further, in a P2P network, any single arbitrary chosen entity can be removed from the network without having the network suffering from any loss of network service. Now, in light of the idea outlined by \cite{Schollmeier_2001}, a \emph{P2P energy network} can be defined as follows:
\begin{definition}
A P2P energy network can be defined as a network, in which the members of the network can share a part of their resources (for example, renewable energy and storage space) and information to attain certain energy-related objectives. Example of such objectives include renewable energy usage maximization, electricity cost reduction, peak load shaving, and network operation and investment cost minimization. Each member can be a provider, a receiver or both of the network resources and can directly communicate with rest of the peers of the network without any intervention from third party controller.  Further, a new peer can be added to or an old peer can be removed from the network without altering the operational structure of the system.
\label{def:definition1}
\end{definition}
\begin{figure}
\centering
\includegraphics[width=0.6\columnwidth]{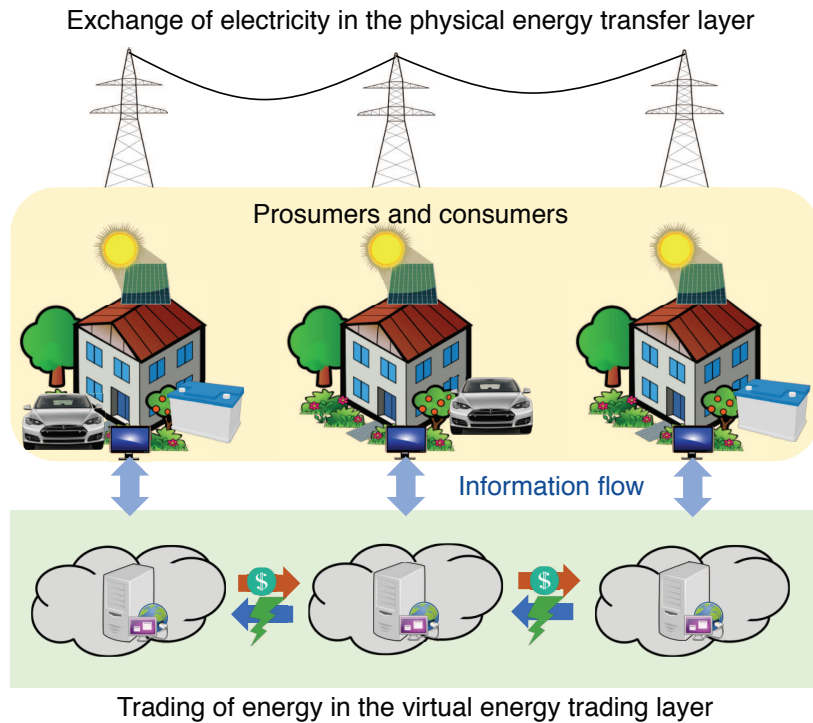}
\caption{Demonstration of different layers in a typical P2P energy network.}
\label{fig:P2PNetwork}
\end{figure}

As shown in Fig.~\ref{fig:P2PNetwork}, a P2P energy network generally consists of two layers~\cite{Mengelkamp_AppliedEnergy_2017}: 1) a virtual energy trading layer and 2) a physical energy transfer layer. The \emph{virtual energy trading layer} essentially resembles a local electricity market, in which participating peers of the network exchange necessary information to decide on the resource type, resource amount, and price per unit of resource exchanged with one another. A virtual layer needs to be based on a secured information system (for example, blockchain based architecture~\cite{PowerLedger_2017}) that provides equal access to each peer. In the virtual layer, the generation, demand, and consumption data of a peer are transferred from its smart meter over a secured communication network based on which buy and sell orders are created. Then, using the created orders, an appropriate market mechanism is used to facilitate energy trading. Once all buy and sell orders from different peers are matched, the payment is carried out and the exchange of energy takes place over the physical layer.

The \emph{physical energy transfer layer}, on the other hand, is essentially a distribution grid, which is used for the physical transfer of electricity between peers. This physical network can either be a traditional distribution network, which is provided and maintained by an independent system operator (ISO) or be a separate physical microgrid distribution grid in conjunction with the traditional grid. It is worth noting that the financial transaction between the peers in the virtual layer does not directly control the physical delivery of electricity. In fact, the payment from a buyer to a seller is done to initiate the transfer of seller's renewable energy into the distribution grid.

Now, based on how different prosumers interact with one another over a P2P network for energy, the existing P2P energy trading literature can be divided into three general categories. 
\subsubsection{Microgrid-to-Microgrid energy trading}P2P energy trading mechanisms that fall into this category mainly focus on handling the mismatch between the supply and demand of energy within microgrids by improving the utilization of renewable energy resources. Such mechanisms, on the one hand, reduce electricity bills for the participating microgrids in the P2P energy trading~\cite{Liu_Smartgridcomm_2015}. It is possible to smooth power generations at the microgrid by regulating the power flow within the grid despite the stochastic load demand~\cite{Fathi_TSE_2013}. Further, incorporating the aspect of security within P2P trading allows the proliferation of distributed energy resources and flexible demands to realize a more resilient grid in both normal and emergency conditions~\cite{Moslehi_TPS_2018}. One example of this type of study can be found in  \cite{Wang_TIE_2016}.
\subsubsection{Intra-microgrid P2P energy trading}This category of energy trading techniques models the decision making process of exchanging energy between prosumers within a microgrid. For instance, the authors in \cite{Morstyn_TPS_2018} introduces a new concept of P2P trading based on multi-class energy management. Here, energy is treated as a heterogeneous product, based on attributes of its source which are perceived by prosumers to have value, such as generation technology, location in the network and the owner's reputation. Existing studies under this category can further be divided into two kinds. The first kind of studies deal with financial allocation based on renewable energy generation. For example, in \cite{Sorin_IEEE_2018}, the authors propose a consensus based approach to P2P trading with product differentiation. \cite{Liu_TPS_2017} proposes an energy sharing model with price based demand response for peer-to-peer network within a microgrid. In \cite{Nguyen_AE_Oct_2018}, the authors propose an optimization model for battery systems in a P2P energy trading market to optimize the use of rooftop solar generation. \cite{Chapman_IEEE_2018} proposes a methodology based on sensitivity analysis to assess the impact of P2P transactions on the network and to guarantee an exchange of energy that does not violate network constraints. A similar kind of P2P energy trading between smart homes is proposed in \cite{Alam_AE_Mar_2019}. The second type of studies, on the other hand, focus on incentivizing prosumers for coordinated scheduling of flexible loads and energy storage. Examples of such studies can be found in \cite{Morstyn_TSG_2018}, \cite{Paudel_IEEE_2018} and \cite{Han_IEEE_2019}. In \cite{Morstyn_TSG_2018}, the authors introduce real-time and forward markets, consisting of energy contracts offered between generators with fuel-based sources, for incentivizing coordination between the owners of large-scale and small-scale energy resources at different levels of the power system. An evolutionary game-theoretic approach and a Stackelberg game are used in \cite{Paudel_IEEE_2018} for modeling the dynamics of buyers and seller to capture their interactions in a P2P energy network. Finally, \cite{Han_IEEE_2019} develops a cooperative game theoretic approach for incentivizing prosumers to form coalitions for P2P trading. A detail overview of different P2P energy market models can also be found in \cite{Sousa_RSER_April_2019}.
\subsubsection{Peer to microgrid energy trading}An interesting concept of energy exchange between peer and microgrid is studied in \cite{Stevanoni_TSG_2018}. The objective is the long-term planning of the peer connected industrial microgrids by coupling the decision of long-term investments and short-term operations via two game theoretical frameworks that allow the modeling of the different, even conflicting, objectives of the stakeholders. The developed tool is tested on a virtual industrial microgrid to present the technical and economic outputs.  Other than above-mentioned categories, which are summarized in Table~\ref{table_P2PPreliminaries}, the application of P2P trading is also found in managing energy for electric vehicles, e.g., see \cite{Kang_TII_2017} and \cite{Alvaro-Hermana_ITSM_2016}. 

Based on the above discussion, clearly, the interaction and cooperation between peers within energy networks are critical for the successful conductance of P2P trading. Therefore, it is critical that the prosumers participate in the trading process. However, to enable this participation, P2P energy trading mechanisms need to possess some qualities that motivate the prosumers to engage in the trading. To this end, how a motivational psychology framework can be used to motivate prosumers to participate in P2P trading is discussed in the following section.
\begin{table*}[t]
\centering
\caption{Summary of different types of P2P energy trading scheme available in the literature.}
\includegraphics[width=\textwidth]{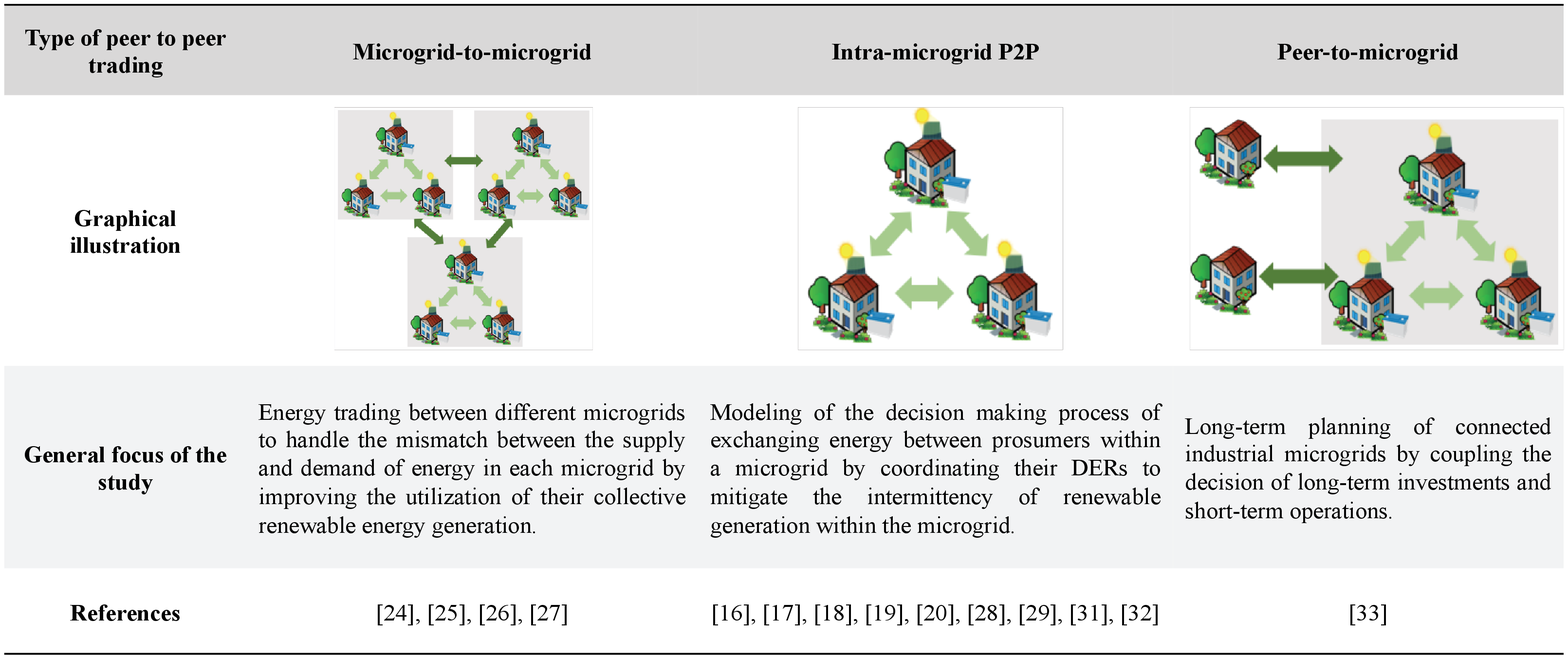}
\label{table_P2PPreliminaries}
\end{table*}
\subsection{Motivational Psychology}\label{sec:motivationalpsychology} Motivation is a cognitive process that establishes reasons for human behavior~\cite{Hockenbury:2003}. Motivational psychology is a branch of behavioral science that studies the psychological process, which regulates a human's behavior based on his perspective, belief, and opinion towards an action~\cite{Beebe:1999}. For example, according to \cite{Lohse_SIGCHI:1998}, the design of a user interface can significantly impact human psychology and subsequently affect the trading of goods in a cyberstore. The application of motivational psychology can be found in public health~\cite{Glanz_ARPH:2010}, education system~\cite{Hermida_TNE:2015}, human service~\cite{Aryee_JM:2016}, economics~\cite{Stein_IO:2017}, medicine~\cite{Carr_IO:2017}, engineering~\cite{Franca_IST:2014}, and energy management~\cite{Tushar_TCSS_2018}. 

\subsubsection{Change of human behavior}According to the \emph{stages of change model}~\cite{Helen_CHI:2010}, the change in human behavior towards taking an action, for example, in this case, participating in P2P trading, does not change abruptly. Rather, a person needs to go through a series of multiple behavioral stages.  In stage 1, a person is unfamiliar with the potential benefit of participating in P2P energy trading. Subsequently, they are unwilling to change their behavior towards participating in P2P trading in the foreseeable future, e.g., six months~\cite{JO_AJHP:1995}. In stage 2, the person becomes aware of the benefit of participating in P2P trading and acknowledges the potential drawback that may arise from not participating. He becomes interested to learn more about P2P trading and its potential economic and environmental benefits. Nonetheless, he may still be not committed to actual participation~\cite{Miller:2002}. 

\begin{figure}[t]
\centering
\includegraphics[width=0.6\columnwidth]{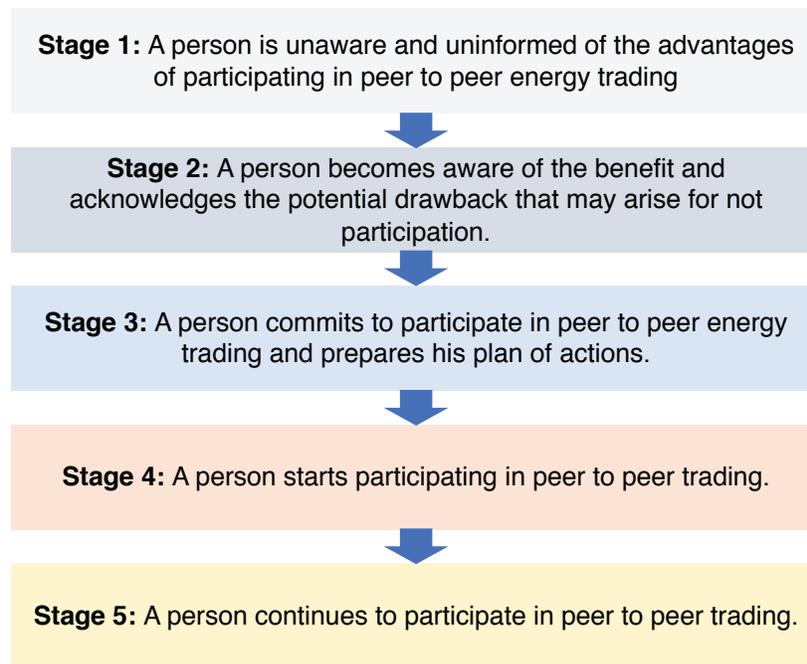}
\caption{Summary of different stages that a prosumer needs to pass through before agreeing to continue participating in P2P trading.}
\label{fig:Stages}
\end{figure}

In stage 3, the person becomes convinced of the potential environmental, social, and economic benefit that P2P trading can bring and \emph{becomes ready} to participate in P2P energy trading and plans to take necessary actions~\cite{Miller:2002}.  In stage 4, the person starts participating in P2P trading~\cite{JO_AJHP:1995} for the first time and thus becomes a part of sustainable energy practice. Finally, when he reaches the stage 5, he continues working towards sustaining his behavioral change~\cite{JO_AJHP:1995}. In other words, the individual continuously participates in P2P trading with other prosumers of the network. An overview of these different motivational stages is shown in Fig.~\ref{fig:Stages}.

However, a key challenge is to enable an individual (or prosumer) to pass through these motivational stages in order to ensure his sustainable participation in P2P trading. As such, a number of motivational models are proposed in the literature as detailed in the next section.
\subsubsection{Models to motivate prosumers towards sustainability}Based on the discussion in the previous section, clearly, a successful P2P trading platform requires a large number of \emph{Stage 5} prosumers. From this perspective, there are a number of motivational psychology models that can be used to study how to influence prosumers to continue their participation in P2P trading. Examples of such models include attitude model, rational-economic model, information model, elaboration likelihood model, and positive reinforcement model.
\paragraph{Attitude model}According to this model, the environmental friendly behavior of a person is automatically influenced by his favorable attitude towards the environment~\cite{Shipworth:2000}. In other words, if a prosumer is convinced of the fact that participating in P2P trading will benefit the environment, he will participate. Nevertheless, such positive attitude can also be affected by the user's view on privacy~\cite{Geert_EP:2013} and his sense of contribution towards environmental sustainability~\cite{Gabriell_CHB:2014}.
\paragraph{Rational-economic model}This model establishes that economic benefit significantly affects people's pro-environmental decision~\cite{Shipworth:2000}. In other words, monetary cost and revenue resulting from participating in P2P trading can effectively motivate people to be responsible and take necessary actions to participate in P2P trading.
\paragraph{Information model} According to the information model, if a person is informed about a problem and the implication of the solution, he would come forward and take appropriate measures to find a solution~\cite{Shipworth:2000}. Moreover, information on other relevant aspects of the issue such as privacy, environmental sustainability, cost, and revenue can also motivate an individual to be proactive about engaging in taking necessary action~\cite{Withanage_JME:2016}.
\paragraph{Elaboration likelihood model}According to this model, there are two paths of communicating with and motivating people: the central path and the peripheral path~\cite{Petty_1986}. The central path is assumed to be appropriate for motivation when an individual cares about the issue and has access to the information with a minimum of distraction. However, he may potentially move away from the advocated position if the subject rehearses unfavorable thoughts due to the ambiguity of the message. In this case, the peripheral path is more suitable that tries to associate the advocated position with things the receiver already thinks positively towards (e.g., monetary benefit), using an expert appeal.
\paragraph{Positive reinforcement model}A positive reinforcement essentially refers to the case when an individual's response to a situation is followed by a reinforcing stimulus that increases the possibility of having the same response from him when a similar situation occurs~\cite{Hockenbury:2003}. For example, receiving a considerable monetary benefit by participating in P2P energy trading can positively reinforce a user to participate again in the future.
\begin{figure}[t]
\centering
\includegraphics[width=0.6\columnwidth]{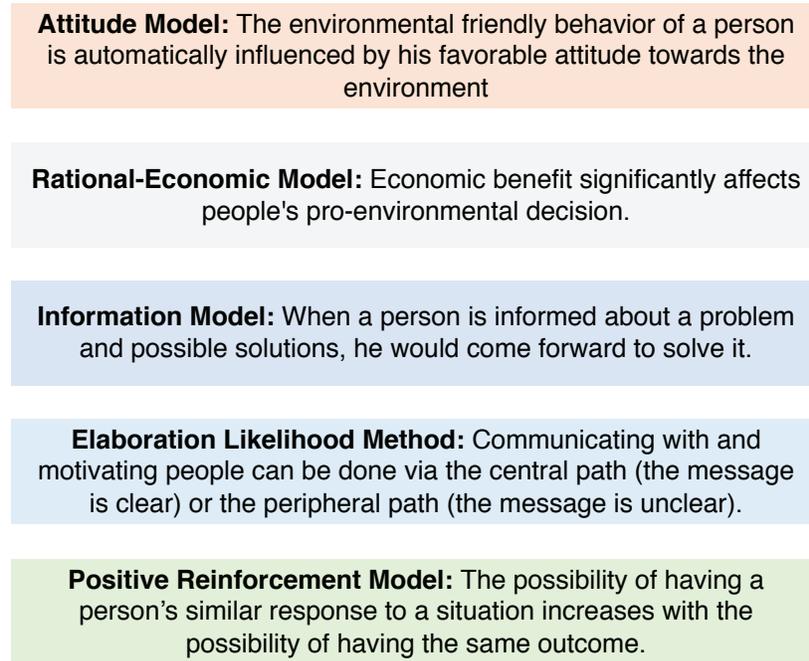}
\caption{Summary of the motivational models that needs to be supported by the P2P energy trading scheme to encourage widespread participation of prosumers.}
\label{fig:Models}
\end{figure}

Now, based on the above discussion and their summaries in Fig.~\ref{fig:Stages} and Fig~\ref{fig:Models}, it is reasonable to say that it could be possible to enable widespread participation of prosumers in P2P trading if the trading mechanism can be developed in such a way that the outcome of participation satisfies the relevant motivational models. From this perspective, what follows is an overview of  a P2P energy trading technique, which is developed to support the motivational psychology models outlined in this section.
\section{Overview of A Motivational Psychology Supported P2P Energy Trading}\label{sec:gamedesign}
\begin{figure}[t]
\centering
\includegraphics[width=0.6\columnwidth]{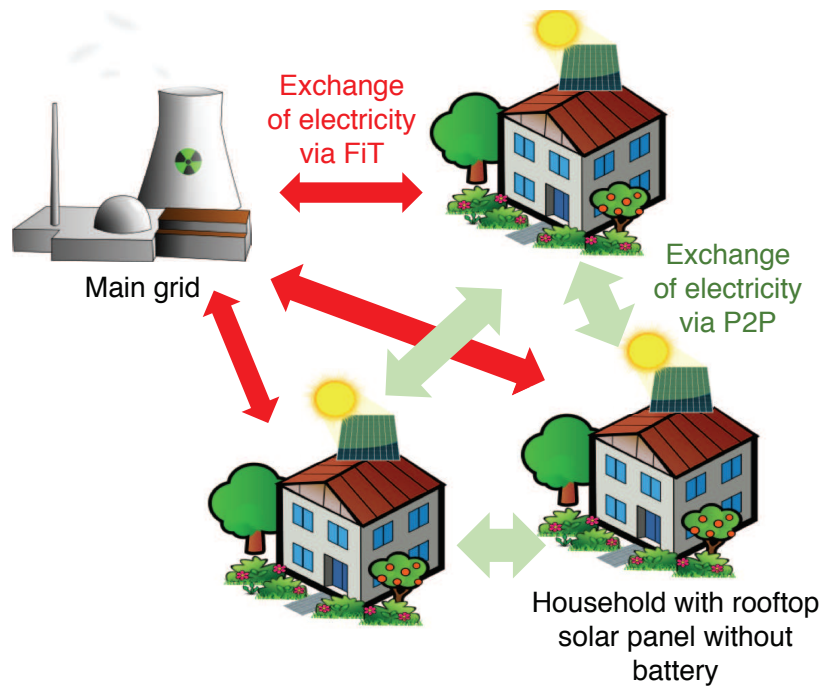}
\caption{This figure illustrates an example scenario in which a number of households participate in the P2P energy trading and/or the FiT scheme.}
\label{fig:P2P}
\end{figure}

To discuss how to develop a P2P energy trading technique whose outcome may satisfy the motivational psychology models, consider a very simple scenario, in which a number of households in a community are connected to the grid via grid-connected solar system without battery~\cite{Tushar_TSG_2017}. In general, these households participate in energy trading with the grid through a FiT scheme~\cite{Liang-ChengYe-AE:2017}. However, once the households are connected to a P2P network, they begin to interact with one another as peers of the network to exchange electricity~\cite{Mengelkamp_AppliedEnergy_2017}. Nonetheless, each household may also interact with the grid if there is an energy surplus or deficiency after completing P2P trading. An illustration of the FiT and the P2P energy trading scheme is shown in Fig.~\ref{fig:P2P}. Now, due to the interactive nature of the energy trading process, a game theoretic approach is used to develop the P2P trading scheme in this section.
\subsection{Brief introduction to game theory}Game theory is a mathematical tool that analyzes strategies of competitive situations where the outcome of a player's action depends on the actions of other players~\cite{GameTheoryBook_1998}. Game theory can be broadly divided into two types: 1) non-cooperative game and 2) cooperative game.
\subsubsection{Non-cooperative game}A non-cooperative game deals with the strategic decision-making process of a number of independent players that have totally or partially conflicting interest over the outcome of a decision that is affected by their choice of actions. Now, if the players act only once, either simultaneously or at different points in time, the game is called a static non-cooperative game. In a dynamic game, on the other hand, players act more than once and have more information regarding the choice of other players. Further, time plays a central role in the decision-making process of each player in a dynamic game.

The most popular solution concept of the non-cooperative game is the Nash equilibrium. Essentially, the Nash equilibrium can be defined as a vector of actions that leads a non-cooperative game to reach a stable state, in which no player can be better paid off by unilaterally deviating from its Nash equilibrium action when the actions of other players are according to the Nash equilibrium. 
\subsubsection{Cooperative game} A cooperative game focuses on how one can provide incentives to independent decision makers so that they act together as one entity in order to improve their position (or utility) in the game. In particular, it studies the terms and conditions under which a number of players may agree to form a coalition (Nash bargaining) and the formation of coalitions (coalition game)~\cite{Tushar_SPM_2018}. There are three categories of coalition games.
\begin{enumerate}
\item\emph{Canonical coalition game} that studies the properties and stability of the grand coalition, the gains resulting from the coalition, and how to distribute these gains in a fair manner to the players. The most renowned solution concept for the coalitional game is the core, which is directly related to the stability of the grand coalition.
\item\emph{Coalition formation game}, which studies the formation of a coalition structure through players' interaction, and then studies the properties of the structure and its adaptability to environmental variations.
\item\emph{Coalition graph game} deals with such connectivity of communications between players with the main objectives to derive low complexity distributed algorithms for players that wish to build a network graph and to study the properties such as stability and efficiency of the formed network graph.
\end{enumerate}
\subsection{P2P energy trading scheme}To design a P2P energy trading scheme, the framework of a canonical coalition game is used. The choice of this game is motivated by the following factor: P2P trading necessitates the prosumers to rely on each other for trading electricity. Thus, it can be seen as a group of households working together to achieve both global (e.g., reducing CO$_2$ emission) and local (e.g., reducing the cost of electricity) objectives. As such, by studying the properties and stability of such a group, it would be possible to comment on whether the participating households can form a stable coalition with each other, which subsequently answers the question of whether the household finds it beneficial for them to remain in the coalition in order to participate in P2P trading. 
\subsubsection{Stability of a coalition}The stability of a canonical coalition game refers to the fact that the grand coalition\footnote{The coalition of all prosumers in the network.} is stable. That means once the grand coalition of prosumers is formed for P2P trading, no prosumer will have an incentive to leave the grand coalition for further improving its benefit. Mathematically, a grand coalition is stable if 1) it has a non-empty core, which is a vector of revenue allocation that ensures that each prosumer that forms a coalition with other prosumers for trading energy receives a revenue such that it has no incentive to leave the grand coalition~\cite{Saad-coalition_2009}, and 2) the revenue that each prosumer will achieve for participating in P2P trading always lie within the core~\cite{Saad-coalition_2009}. 
\begin{definition}\label{definition:1}
From a coalition game perspective, peer-to-peer energy trading between the prosumers will be stable if:
\begin{itemize}
\item The game possesses a non-empty core, that is, the value function of the coalition needs to be superadditive. For the definition of core and superadditivity, please see \cite{Saad-coalition_2009}. 
\item The P2P energy trading price needs to ensure that the revenue that each prosumer may attain lies within the core \cite{Saad-coalition_2009}. In other words, the P2P trading price needs to always encourage the prosumers to participate.
\end{itemize}
\end{definition}
\subsubsection{Development of the trading scheme}To use a canonical coalition game with the above properties for designing a P2P trading scheme, it is assumed that a number of prosumers with rooftop solar panels (without batteries) are connected to a central power station (CPS). They are also connected to each other through a secure information system via their smart meters for all necessary communication and transactions of P2P trading. At any particular time, each prosumer $n$ has a particular amount of electricity surplus $E_{n,s}$ from the supply from its solar panel and deficiency $E_{n,d}$. If the prosumer has any surplus, i.e., $E_{n,s}>0$,  it sells it to meet the demand of other peers as the first priority and then to  the CPS, if needed. Similarly, if there is any electricity deficiency, i.e., $E_{n,d}>0$, a prosumer first buys the necessary amount from the peers within the network and then from the CPS, if required. Thus, in a P2P trading, all the prosumers cooperate with one another to trade the surplus energy among themselves as a first priority, and then interact with the grid, if necessary, to sell or buy the total excess or deficient energy respectively.

Now, to enable such a trading of electricity, a canonical coalition game $\Gamma = \left(\mathcal{N}_s\cup\mathcal{N}_b,\nu\right)$ is considered. Here, $\mathcal{N}_s$ and $\mathcal{N}_b$ are the sets of sellers and buyers of electricity respectively within the P2P network and $\nu$ is the value of coalition~\cite{Saad-coalition_2009}, which, in this case, refers to the monetary benefit that prosumers, as a coalition group, may achieve for participating in P2P trading. The monetary benefit $\nu$ to the group of prosumers can be viewed as the difference between revenue and cost, i.e., $\nu=p_{g,b}[E_s]^{+}-p_{g,s}[E_d]^{+}$, where $E_s = \sum_{n\in\mathcal{N}_s}E_{n,s}-\sum_{n\in\mathcal{N}_b}E_{n,d}$, $E_d = -E_s$, $[\cdot]^{+} = \max(0,\cdot)$, and $p_{g,s}$ and $p_{g,b}$ are respectively selling and buying prices by the grid. Indeed, the value function $\nu$ is a concave function and superadditive. 

Now, note that $p_{g,s}>p_{g,b}$~\cite{Tushar_TIE_2015}, which is due to the fact that retailers pay prosumers less for their excess generation, since it is considered non-dispatchable. Therefore, a suitable P2P trading price can potentially motivate the prosumers to cooperate with one another in order to trade electricity among themselves, if possible, instead of trading with the CPS. To this end, the mid market rate~\cite{Mid_Market_2017} is proposed to use as the pricing mechanism for this coalition game based P2P trading.
\begin{figure}[t]
\centering
\includegraphics[width=0.6\columnwidth]{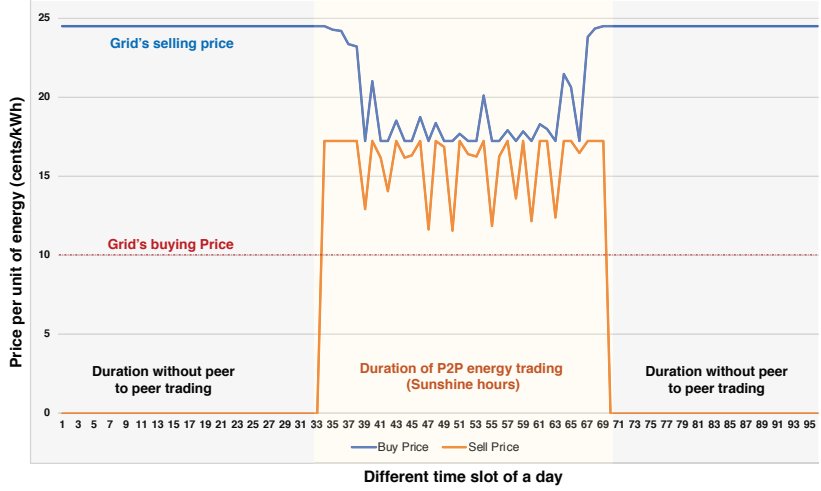}
\caption{This figure illustrates how the mid market rate pricing scheme sets the P2P buying and selling price always within the range $\left[p_{g,b}, p_{g,s}\right]$. The value of $p_{g,b}$ and $p_{g,s}$ are chosen according to the electricity price in Brisbane, Australia.}
\label{fig:Core}
\end{figure}

\subsubsection{Setting trading price through mid-market rate}In mid-market rate, price of energy trading is set based on three scenarios~\cite{Mid_Market_2017}. They are explained as follows.
\paragraph{Scenario 1 ($E_s =0, E_d = 0$)}This scenario refers to the case when the total available surplus energy of the prosumers is equal to their demand. In other words, prosumers, if participate in P2P trading, do not need to trade any electricity with the CPS. According to mid market rate, the P2P trading price (both selling $p_s$ and buying price $p_b$) is $p_s=p_b=\frac{p_{g,s}+p_{g,b}}{2}$.
\paragraph{Scenario 2 ($E_s > 0$)}This scenario refers to the case when the total available surplus energy of the prosumers is greater than the total demand. In other words, prosumers, if participate in P2P trading, can not only meet the demand of prosumers with energy deficiency, but can also sell a part of it to the CPS. Now, according to mid market rate, the P2P trading price buying price $p_b$ is as same as that in scenario 1. However, due to selling to the CPS, the P2P selling price $p_s$ per unit of electricity becomes $p_s = \frac{p_b\sum_{n\in\mathcal{N}_b}E_{n,d}+p_{g,b}E_s}{\sum_{n\in\mathcal{N}_s}E_{n,s}}$.
\paragraph{Scenario 3 ($E_d > 0$)}This scenario refers to the case when the total available surplus energy of the prosumers is smaller than the total demand. In other words, prosumers, if participate in P2P trading, cannot meet the demand of prosumers with energy deficiency and therefore need to buy excess energy from the CPS. In this scenario, the P2P buying and selling prices become $p_b = \frac{p_s\sum_{n\in\mathcal{N}_s}E_{n,s}+p_{g,s}E_d}{\sum_{n\in\mathcal{N}_b}E_{n,d}}$ and $p_s=\frac{p_{g,s}+p_{g,b}}{2}$ respectively.

Now, based on the above discussion and Definition~\ref{definition:1}, it can be concluded that the developed P2P trading scheme ensures a stable coalition among the participating prosumers due to following reasons.
\begin{itemize}
\item The value function $\nu$ is concave and superadditive. Therefore,  the canonical coalition game has a non-empty core by following~\cite{Lee_JSAC_2014}. 
\item The mid market rate is used for the P2P pricing. Evidently, as shown in Fig.~\ref{fig:Core}, using mid market rate always allows prosumers to buy energy from other prosumers with energy surplus as a price $p_b$ that is lower than the grid's selling price $p_{g,s}$ per unit of electricity. On the other hand, by participating in the P2P trading, prosumers with energy surplus can sell their energy to other prosumers at a price $p_s>p_{g,b}$. This pricing ensures that the revenue distribution that each prosumer achieve for participating in P2P trading lies within the core of the game. 
\end{itemize}
Therefore, prosumers will be interested to cooperate with one another to participate in P2P trading. 
\begin{remark}
Here, it is important to note that the discussion in Table II only emphsizes cost and CO$_2$ reduction, but not other objectives of P2P trading as outlined in Definition~\ref{def:definition1}. This is due to the fact that the study focuses on the benefit to the prosumers in order to encourage them to participate in the trading by satisfying a number of motivational models and these two objectives are relevant for that purpose (other objectives are relevant to the grid). Of course, increasing the use of renewable energy is also important for the prosumers. However, CO$_2$ reduction is the result of using renewable energy and therefore this aspect is not discussed separately.
\end{remark}

Now, how the stable P2P trading scheme satisfies the motivational psychology models explained in Section~\ref{sec:motivationalpsychology} will be discussed in the following section. 

\subsection{How the proposed energy trading scheme can psychologically motivate prosumers}
\begin{figure}[h!]
\centering
\includegraphics[width=0.4\columnwidth]{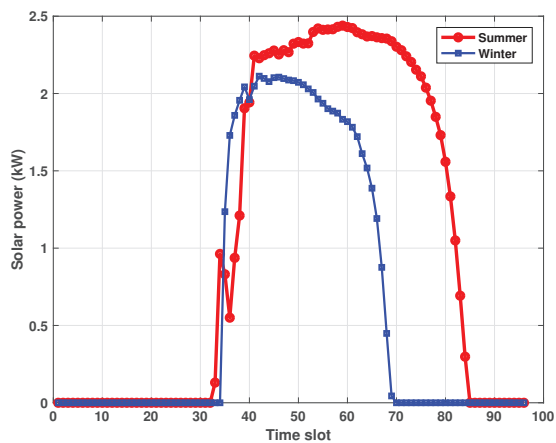}
\caption{This figure shows the solar data used in this paper for simulation.}
\label{fig:SolarData}
\end{figure}
To show how the proposed P2P trading scheme supports the motivational psychology framework, first, we present a number of results based on the game theoretic scheme in this paper. Then, we will show the relevance of the results to different motivational psychology models to demonstrate how the proposed scheme exhibits the properties of attracting users to participate in energy trading. Publicly available real-data on solar generation\footnote{Public solar data is collected from IEEE PES ISS website.} and household energy demand of residential consumers\footnote{Residential data was taken from the website of National Energy Efficiency Alliance (NEEA).} are used in the numerical simulation of the designed P2P trading scheme was used. A demonstration of solar data of a selected day that is used in this paper is shown in Fig.~\ref{fig:SolarData}. For each residential consumer, it was assumed that they were equipped with a 3kWp solar panel and 15 minute resolution data was used. The data used for this case study was recorded in December 2013. The values for the grid's buying price and FiT price were assumed to be 24.6 and 10 \cent/kWh respectively (these values are referenced to the electricity price in Brisbane, Australia as an example). Furthermore, to calculate the production of CO$_2$ from fossil fuel based generation, it was assumed that the power stations are driven by natural gas that produces a carbon footprint of 0.55 Kg per kWh~\cite{Tushar_TCSS_2018}.

The results are compared with an FiT scheme to show the effectiveness of the proposed scheme. The comparison with an FiT scheme is chosen intentionally due to the relevance of FiT with P2P trading in engaging customer with grid-tie solar system without storage to trade their energy. Of course, there are schemes, other than FiT, that can also be used for renewable energy management, such as a grid-tie-solar system with storage. However, since the developed P2P trading scheme assumed a grid-tie system without storage, comparing with a solar system with storage is not valid. Furthermore, although grid-tie-solar systems with storages can be found in households in the recent time, such setups are expensive due to the cost of the battery, procurement of the battery, high battery degradation cost, and has a very long payback period \cite{CSIRO_2015}. Therefore, there is unlikely to have extensive uptakes of battery storage by households in the near future.

\begin{table*}[h]
\centering
\caption{This table demonstrates how much prosumers can save, compared to an FiT scheme, by participating in P2P energy trading.}
\small
\begin{tabular}{|p{2cm}|p{4cm}|p{4cm}|}
\hline
\textbf{Prosumer} & \textbf{Cost savings in Summer (\cent)} & \textbf{Cost savings in Winter (\cent)}\\
\hline
1 & 63.55 & 42.61 \\
\hline
2 & 61.19 & 34.25 \\
\hline
3 & 100.78 & 63.92 \\
\hline
4 & 118.73 & 86.96\\
\hline
5 & 101.23  & 61.50\\
\hline
\end{tabular}
\label{table:costsavings}
\end{table*}

In Table \ref{table:costsavings}, we consider five prosumers participating in P2P trading and show how much they can save by taking part in P2P trading with one another. In particular, we show how much each prosumer can save compared to an FiT scheme for participating in P2P trading in summer and winter. As shown in Table~\ref{table:costsavings}, by participating in the proposed P2P trading scheme, prosumers can always save money compared to what they could save by trading via the FiT scheme. Since the total sunshine duration is longer and the intensity of sunlight is higher in Summer compared to Winter, the amount of solar production is also higher in Summer. Consequently, the prosumers can save more in Summer than Winter.

\begin{figure}[h]
\centering
\includegraphics[width=0.45\columnwidth]{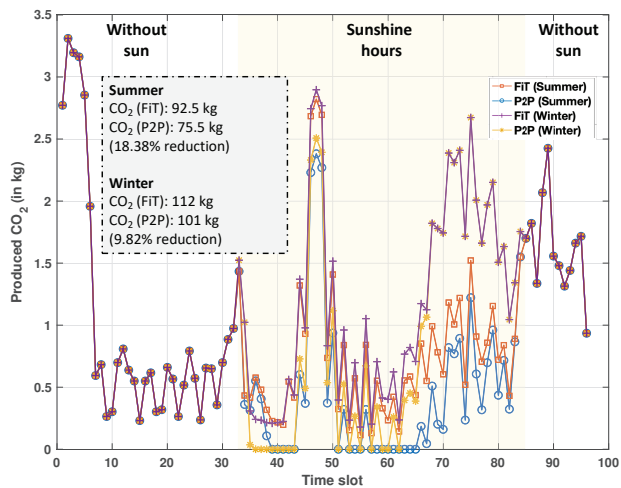}
\caption{Demonstration of the amount of CO$_2$ production in a single day.}
\label{fig:CO2Production}
\end{figure}
The variation in sunshine hours and the sunlight intensity between Summer and Winter also affect the production of CO$_2$ in these two seasons, which is shown in Fig.~\ref{fig:CO2Production}. Of course, under the P2P and FiT scheme the total amount of renewable generation is the same. However, under the P2P scheme, prosumers are directly aware of the amount of renewable energy they are obtaining from other prosumers, whereas, under the FiT scheme, each prosumer is only aware of their own individual renewable generation. This affects the production of CO$_2$ within the selected community as shown in Fig.~\ref{fig:CO2Production}. According to Fig.~\ref{fig:CO2Production},
\begin{itemize}
\item The generation of CO$_2$ for both P2P and FiT schemes are higher in Winter compared to that in Summer. This is mainly due to the longer sunshine hours and higher solar intensity in Summer.
\item During the times outside sunshine hours, the production of CO$_2$ is same in both P2P and FiT schemes. Since, no storage is considered in the system, the proposed P2P scheme cannot perform better than the FiT scheme when there is no sun. During sunshine hours, however, P2P shows considerable performance improvement in terms of CO$_2$ production compared to the FiT.
\item Energy trading via the proposed P2P trading technique exhibits lower amount of CO$_2$ production in both Summer and Winter, compared to those for FiT scheme. This is due to the fact that in the FiT scheme, as considered in this study, a prosumer buys its deficient energy from the grid and the grid burns fossil\footnote{It is assumed that the grid does not use any renewable energy to meet the demand of prosumers.} fuel to produce that energy. Under P2P trading, the deficient energy of any prosumer is first met by the surplus energy from another peer of the network and a prosumer only consume from the grid if there is no surplus left in the overall peer-to-peer network. Therefore, the demand to the grid is lower in P2P scheme, which subsequently results in lower CO$_2$ production. For example, in Summer and Winter, the proposed P2P scheme demonstrates 18.38\% and 9.82\% reduction in CO$_2$ production respectively in comparison to the FiT scheme.
\end{itemize}
\begin{table*}[h]
\centering
\caption{This table demonstrate the environmental and economic benefits of the developed P2P energy trading for different days in a week.}
\includegraphics[width=\textwidth]{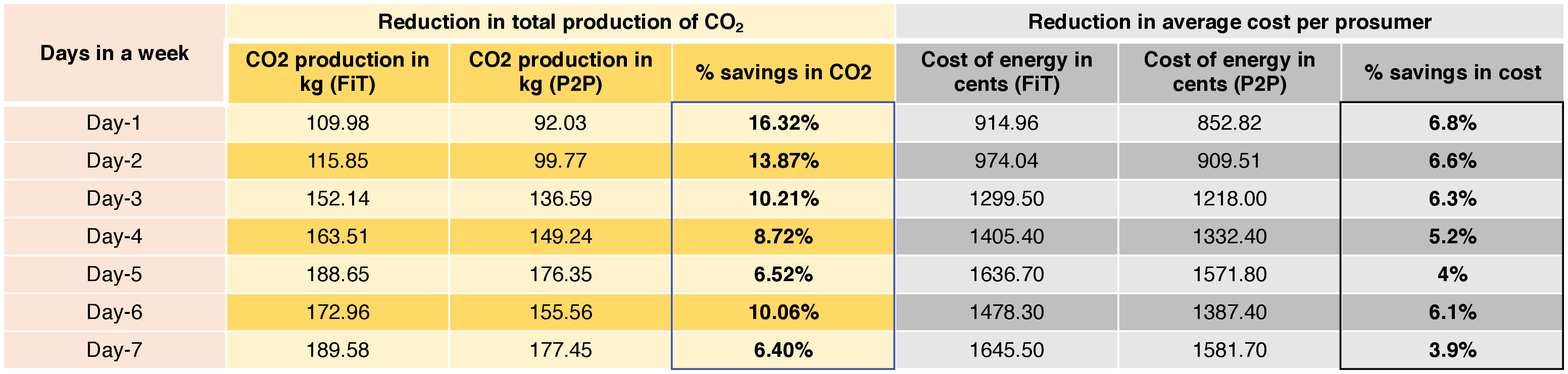}
\label{tab:properties}
\end{table*}
It is important to note such benefits of reduction in cost and CO$_2$ production of the P2P scheme over the FiT scheme always hold true. For example, in Table~\ref{tab:properties}, we show the average reduction in CO$_2$ production and cost of energy per prosumer for seven consecutive days in Winter. We do not show this result for the Summer, as Summer demonstrates even a better performance. According to this table, whenever prosumers choose to trade their energy by using the proposed P2P trading, it benefits them, in terms of average cost reduction as well as the alleviation of average CO$_2$ production. Thus, based on the results from Table~\ref{table:costsavings}, Fig.~\ref{fig:CO2Production} and Table~\ref{tab:properties}, the proposed P2P energy trading model satisfies all the discussed motivational models as explained below.

\begin{itemize}
\item\emph{Attitude model:} The outcomes of the P2P trading supports the attitude model. This is due to the fact that the designed scheme demonstrates a consistent performance in reducing CO$_2$ production (Fig.~\ref{fig:CO2Production} and Table~\ref{tab:properties}), which has the potential to convinced prosumers who are environmentally friendly to participate in P2P trading. Thus, the P2P trading platform can help make prosumers aware of the amount of CO$_2$ reduction their PV panels are helping to provide. This outcome can motivate prosumers of all motivational stages.
\item\emph{Rational economic model:} In Table~\ref{tab:properties}, a clear benefit of P2P trading in reducing average cost of energy per prosumer is demonstrated, which are $6.8\%$, $6.6\%$, $6.3\%$, $5.2\%$, $4\%$, $6.1\%$, and $3.9\%$ reduction compared to a FiT scheme at different different days of the selected. Similar results are also shown in Table~\ref{table:costsavings}. Consequently, the proposed P2P energy trading scheme satisfies the rational economic model and can be helpful for prosumers at Stage 2, Stage 3, and Stage 4.
\item\emph{Information model:} By providing the prosumers with useful information of the environmental benefit and economical profit that can be obtained by P2P trading, they can be encouraged to participate, which subsequently complies with the information model. This information can helpful to engage prosumers who are at Stage 1 and Stage 2.
\item\emph{Elaboration likelihood model:} While the technical detail of how the P2P trading can be done at household levels is difficult to convey to general prosumers, the benefits such as reduction of electricity bills and environmental sustainability can be easily exploited to convince prosumers of the importance of their participation in the trading process. Note that these are some benefits that prosumers really care about and thus such communication following a peripheral path can help users to be an integral part of P2P trading. This is relevant to the Stage 2 prosumers.
\item\emph{Positive reinforcement model:} A brief look at the performance of the developed P2P energy trading scheme as demonstrated in Table~\ref{tab:properties} emphaizes that its performance is consistent and it continuously help prosumers to reduce their electricity cost and CO$_2$ emission whenever they participate in P2P trading. Thus, based on the definition explained in Section~\ref{sec:motivationalpsychology}, the proposed trading scheme satisfies the positive reinforcement model and help prosumers to reach and continue as Stage 5 participants of the P2P trading.
\end{itemize}
Accordingly, based on the above discussion, which is also summarized in Table~\ref{table.ref1}, it is reasonable to assume that the proposed canonical coalition based P2P trading scheme has the potential to enable the extensive participation of prosumers in P2P trading. In fact, as the participation of prosumers increases, the overall benefits also increase. 
\begin{table*}[t]
\centering
\caption{Summary of how the developed canonical coalition game based peer-to-peer energy trading model satisfy motivational psychology models. A brief discussion of the models can be found in Section~\ref{sec:motivationalpsychology}.}
\small
\begin{tabular}{|m{2cm}|m{2cm}|m{8cm}|m{2cm}|m{2cm}|}
\hline
\textbf{Models} & \textbf{Relevant stage of prosumers}& \shortstack{\textbf{How the property holds?}} & Supporting outcomes & \textbf{References}\\
\hline
Attitude model & Stage 1, 2, 3, 4, 5 & The designed scheme demonstrates a consistent performance in reducing CO$_2$ production in each day of the selected week. & Fig.~\ref{fig:CO2Production} \& Table~\ref{tab:properties}& \cite{Geert_EP:2013,Tushar_TCSS_2018}\\
\hline
Rational economic model & Stage 2, 3, 4  & A clear benefit of peer-to-peer trading in reducing average cost of energy per prosumer is demonstrated, which are $6.8\%$, $6.6\%$, $6.3\%$, $5.2\%$, $4\%$, $6.1\%$, and $3.9\%$ reduction compared to a FiT scheme at different different days of the selected.& Table~\ref{table:costsavings} \& Table~\ref{tab:properties}&\cite{Shipworth:2000,Tushar_TCSS_2018}\\
\hline
Information model & Stage 1, 2  & It provides the prosumers with useful information of the environmental benefit and economical profit that can be obtained by peer-to-peer trading, they can be encouraged to participate. &Table~\ref{table:costsavings}, Table~\ref{tab:properties} \& Fig.~\ref{fig:CO2Production} &\cite{Shipworth:2000,Withanage_JME:2016,Tushar_TCSS_2018}\\
\hline
Elaboration likelihood model & Stage 2  &The proposed scheme uses a peripheral path to help users to understand the benefits of peer-to-peer trading. &Table~\ref{table:costsavings}, Table~\ref{tab:properties} \& Fig.~\ref{fig:CO2Production} &\cite{Petty_1986}\\
\hline
Positive reinforcement & Stage 5 &The performance of the peer-to-peer trading model is consistent, which continuously helps prosumers to reduce their electricity cost and CO$_2$ emission whenever they participate in the peer-to-peer trading. &Table~\ref{tab:properties} &\cite{Hockenbury:2003,Tushar_TCSS_2018}\\
\hline
\end{tabular}
\label{table.ref1}
\end{table*}

For example, in Table \ref{table:Effect_of_prosumers}, we show how the number of prosumers that participate in the P2P trading affects the total reduction in energy cost and CO$_2$ production within the network. To do so, we increase the number of prosumers in the considered system from 5 to 25, calculate the reduction in total cost and CO$_2$ production compared to the FiT scheme, and then list in Table~\ref{table:Effect_of_prosumers}. As can be seen in the table, while the trend of change in the cost and CO$_2$ production across different seasons remain as same as the results previously shown in Table~\ref{table:costsavings} and Fig.~\ref{fig:CO2Production}, the increase in the number of participating prosumers in the energy trading clearly improve the performance of the overall system. For instance, as the number of prosumers within the network increases from $5$ to $25$, we observe a steady increase in the reduction in total cost from \$4.45 to \$26.07 in Summer and from \$2.89 to \$15.63 in Winter. Similar steady increase in total reduction in CO$_2$ production, compared to the FiT scheme, is also demonstrated for both seasons.

\begin{table*}[h]
\centering
\caption{This table demonstrates how much prosumers can save, compared to an FiT scheme, by participating in P2P energy trading.}
\small
\begin{tabular}{|p{2cm}|p{3cm}|p{3cm}|p{3cm}|p{3cm}|}
\hline
\textbf{Total prosumer} & \textbf{Total cost savings in \$ (Summer)} & \textbf{Cost savings in \$ (Winter)} & \textbf{Total CO$_2$ savings in kg (Summer)} &  \textbf{Total CO$_2$ savings in kg (Winter)}\\
\hline
5 & 4.45 & 2.89 & 16.88 & 10.96\\
\hline
10 & 8.85 & 6.06 & 33.54 & 23\\
\hline
15 & 17.59 & 10.6 & 66.69 & 40.18\\
\hline
20 & 22.27 & 13.34 & 84.21 & 50.60\\
\hline
25 & 26.07 & 15.63 & 98.81 & 59.24\\
\hline
\end{tabular}
\label{table:Effect_of_prosumers}
\end{table*}
\section{Conclusion}\label{sec:conclusion}In this paper, a motivational psychology framework has been introduced for peer-to-peer energy trading with an objective to improve the participation of prosumers. First, preliminaries of peer-to-peer trading have been provided via suitable classification of the state-of-the-art literature. Then, a motivational psychology framework has been introduced detailing different motivational stages. Further, it has been demonstrated how a peer-to-peer trading scheme under the framework of motivational psychology can be designed. For this purpose, a canonical coalition game has been used to develop the peer-to-peer trading scheme. Then, with suitable numerical case studies, how the outcomes of the developed peer-to-peer trading scheme satisfy the introduced motivational psychology models has been demonstrated, and thus, its potential to enable widespread participation of prosumers was established.

Based on the proposed study, the following conclusions can be drawn:
\begin{itemize}
\item To ensure a sustainable operation of peer-to-peer trading within an energy system, prosumers need to participate continuously. Hence, motivating them to participate is an important part of the design of a peer-to-peer trading scheme.
\item Motivational psychology provides a number of models, which, with suitable application, has proven to be effective in other domains for motivating people to work towards achieving particular objectives. These models can be exploited to design a peer-to-peer trading scheme in order to ensure prosumers' sustainable participation.
\item One potential way to design peer-to-peer trading with all necessary properties of motivational psychology is to use game theory, which can effectively capture the desired properties of different motivational models. Therefore, game theory should be considered as a strong candidate to model peer-to-peer trading, when the participation of prosumers is a concern.
\item Motivational psychology could be an important factor to be considered while making policy for bringing peer-to-peer energy trading into the existing electricity market. Feed-in tariff failed to show effectiveness in many countries \cite{FiTPolicy_2015}. To avoid a similar circumstance with peer-to-peer trading, motivational psychology could play a critical role for policy design.
\end{itemize}

\section*{Acknowledgement}This work was supported in part by the Advance Queensland Research Fellowship AQRF11016-17RD2, in part by the SUTD-MIT International Design Centre (idc:idc.sutd.edu.sg), in part by NSFC 61750110529, and in part by Engineering and Physical Sciences Research Council (EP/S000887/1).

\end{document}